# COMPARATIVE PERFORMANCE ANALYSIS OF MULTI-DYNAMIC TIME QUANTUM ROUND ROBIN (MDTQRR) ALGORITHM WITH ARRIVAL TIME

H. S. Behera, Rakesh Mohanty, Sabyasachi Sahu, Sourav Kumar Bhoi

*Dept. of Computer Science and Engineering,*
*Veer Surendra Sai University of Technology(VSSUT)*
*Burla, Sambalpur, Odisha, India-768002*

**Abstract**

CPU being considered a primary computer resource, its scheduling is central to operating-system design. A thorough performance evaluation of various scheduling algorithms manifests that Round Robin Algorithm is considered as optimal in time shared environment because the static time is equally shared among the processes. We have proposed an efficient technique in the process scheduling algorithm by using dynamic time quantum in Round Robin. Our approach is based on the calculation of time quantum twice in single round robin cycle. Taking into consideration the arrival time, we implement the algorithm. Experimental analysis shows better performance of this improved algorithm over the Round Robin algorithm and the Shortest Remaining Burst Round Robin algorithm. It minimizes the overall number of context switches, average waiting time and average turn-around time. Consequently the throughput and CPU utilization is better.

## 1. Introduction

Operating System is the interface between the hardware and the user. It controls and coordinates the use of the hardware among various application programs for various users[5]. Modern operating systems have become more complex, they have evolved from a single task to a multitasking environment in which processes run in a concurrent manner[3]. In multitasking and multiprocessing environment the way the processes are assigned to run on the available CPUs is called *scheduling*. The main goal of the scheduling is to maximize the different performance metrics viz. CPU utilization, throughput and to minimize response time, waiting time and turnaround time and the number of context switches[6]. Scheduling is often implemented in diverse real time applications like routing of data packets in computer networking, controlling traffic in airways, roadways and railways, scheduling of league games etc. This assignment is carried out by software known as a *scheduler* and/or *dispatcher*. Operating systems may feature up to 3 distinct types of a *long-term scheduler* a *mid-term or medium-term scheduler* and a *short-term scheduler*. The names suggest the relative frequency with which these functions are performed. In Round Robin (RR) every process has equal priority and is given a time quantum or time slice after which the process is preempted. Although RR gives improved response time and uses shared resources efficiently. Its limitations are larger waiting time, undesirable overhead and larger turnaround time for processes with variable CPU bursts due to use of static time quantum This motivates us to implement RR algorithm with sorted remaining burst time with dynamic time quantum concept. Another concept employed in this algorithm is the use of more than one cycle instead of a single Round Robin.

### 1.1. *Preliminaries*

A program in execution is called a *process*. The processes, waiting to be assigned to a processor are put in a data structure entity called *ready queue*. The time for which a process holds the CPU is known as *burst time*. The time at which a process arrives for execution is its *arrival time*. *Turnaround time* is the amount of time to execute a particular process, while *waiting time* is the amount of time a process has been waiting in the ready queue. Time expired from the submission of a request by the process till its first response is defined as the *response time*. *Scheduler* selects a process from queues in a manner, for its execution such that the load balance is effective. In *non-preemption,* CPU is assigned to a process; it holds the CPU till its execution is completed. But in *preemption*, running process is forced to release the CPU by the newly arrived process. In *time sharing system*, the CPU executes multiple processes by switching among them very fast. The number of times CPU switches from one process to another is called as the number of *context switches.*

### 1.2. Scheduling algorithms

When there are number of processes in the ready queue, the algorithm which decides the order of execution of those processes is called a *scheduling algorithm*. Various well known CPU scheduling algorithms have been developed viz. First Come First Serve algorithm (FCFS), Shortest Job First algorithm (SJF) and Priority scheduling algorithm. All the above algorithms are non-preemptive in nature and are not suitable for time sharing systems. Shortest Remaining Time Next (SRTN) and Round Robin (RR) are preemptive in nature. RR is most suitable for time sharing systems. But its average output parameters (waiting time, turn-around time etc.) are not feasible enough to be employed in real-time systems.

*1.3. Related Work*

Many work has been done related to this. Abielmona[3] on account of his analytical scrutiny of a innumerable number of scheduling algorithms gives a thorough insight into the factors affecting the performance parameters of a scheduling algorithm. RR algorithm gives better responsiveness but worse average turn-around and waiting time. The Proportional Share Scheduling Algorithm proposed by Helmy and Dekdouk[6] combines low overhead of round robin algorithms besides favoring shortest jobs.. The static time quantum which is a limitation of RR was removed by taking dynamic time quantum by Matarneh[4]. Improved variants of the traditional algorithm, SRBRR[2] and RR is compared with our MDTQRR for highlighting its better efficiency. A rule of thumb is also stated that 80% of the CPU bursts should be shorter than the time quantum. The time quantum that was repeatedly adjusted on a run-time basis according to the burst time of the running processes are considered to improve the waiting time, turn-around time and number of context switches.

*1.4. Organization of the paper*

The paper is divided into four sections. Section I gives a brief introduction on the various aspects of the scheduling algorithms, the approach to the current paper and the motivational factors leading to this improvement. Section II presents the materials and methods used, the pseudo code, flowchart and illustration of our newly proposed Multiple Dynamic Time Quantum Round Robin Algorithm (MDTQRR). In section III, an experimental analysis and Result of our algorithm MDTQRR and its comparison with the static RR algorithm and dynamic SRBRR is presented. Conclusion is presented in section IV followed up by the references used. Tables and figures used have been represented by numbers.

**2. Materials and Methods**

In our work, the RR algorithm is improvised by an astute distribution of time quantum of processes, repeatedly over the whole Round Robin cycle. Static time quantum being a limitation of RR algorithm, we have used the concept of dynamic time quantum. Besides, we have supplemented the use of median with upper quartile, as two concepts in one cycle of RR. Implicating that up to the median$^{th}$ process, we use time quantum MTQ (Median Time Quantum) calculated by Median Quartile formula and for the succeeding processes, we use the Upper Quartile formula to calculate the time quantum UTQ (Upper Quartile Time Quantum). This time quantum is used by the remaining processes and this continues up to the execution of all the processes. In succeeding cycles of the round robin, the median and upper quartiles are again calculated taking into consideration the remaining processes.

Formula1 represents the calculation of time quantum by Median Quartile MQ:

$$MQ = \begin{cases} Y_{(N+1)/2} & \text{if N is odd} \\ \frac{1}{2}(Y_{N/2}) + (Y_{1+N/2}) & \text{if N is even} \end{cases} \quad (1)$$

where, Y is the number located in the middle of the group of numbers in ascending order and N is the number of processes. Formula 2 represents the calculation of time quantum by Upper Quartile Q3:

$$UQ = \frac{3}{4}(N+1) \quad \text{where N is the number of processes} \quad (2)$$

So by these formulas we calculate the time quantum in our proposed algorithm in each round.

$$CRITERIA = [\{MTQ*m\} + \{UTQ*(N-m)\}]/N \quad (3)$$

This variable is used for comparison with the 80% criterion.

Some formulae used for calculation are:

$$TAT = CT - AT \quad (4)$$

$$WT = TAT - BT \quad (5)$$

$$CPU\_TIME = \sum BT_i \quad (6)$$

$$TOTAL\_TIME = \{CPU\_TIME + (NCS*CST)\} \quad (7)$$

$$THROUGHPUT = (float)(N/TOTAL\_TIME) \qquad (8)$$

$$CPU\_UTIL = (float)(CPU\_TIME/TOTAL\ TIME) \qquad (9)$$

where  *TAT* is the Turn-Around Time,

*CT* is the Completion Time

*BT* is the Burst Time

*AT* is the Arrival Time

*NCS* is the number of context switches

*CST* is the Context Switch Time

and *THROUGHPUT* and *CPU_UTIL* are the variables for calculating the Throughput and CPU Utilization respectively.

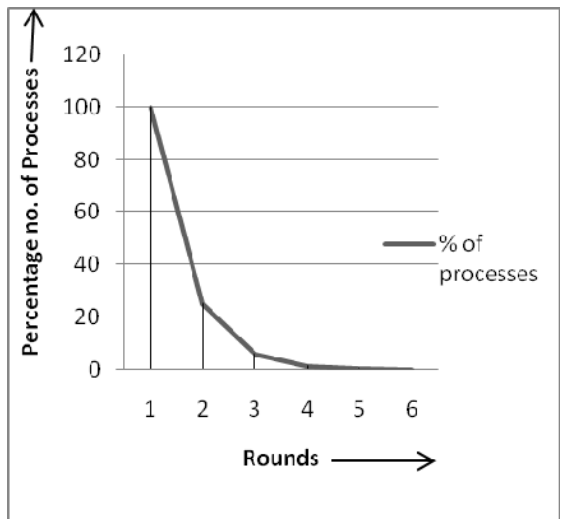

.Fig.1: The rate of decrease in the number of processes in each round.

This means that 75% of the processes will be terminated through the first round and as time quantum is calculated repeatedly for each round then 25% of the remaining processes will be terminated in the consecutive rounds. This process manifests that the maximum number of rounds will be less than or equal to 6 irrespective of the number of process or their burst time. Figure 1 shows the significant decrease of the number of processes in each round. The significant decrease of the number of processes, inevitably will lead to significant reduction in the number of context switch, which may pose high overhead on the operating system in many cases. The number of context switches can be represented mathematically as follows:

$$Q_T = [\sum k_r] - 1 \qquad (10)$$

Where
$Q_T$ = The total number of context switches
$r$ = The total number of rounds, r = 1, 2…6
$k_r$ = The total number of processes in each round

### 3. Our Proposed MDTQRR Algorithm

In our proposed algorithm, the time quantum is taken as the burst time of the median of all the processes. The scheduling continues with the same time quanta up to the median$^{th}$ process. For the succeeding processes, the time quantum is determined by taking the burst time of Upper Quartile of all processes. This whole operation occurs in a single scheduling cycle of the processes sorted in ascending order of the burst time of all the processes.

*3.1. Uniqueness Of Our Approach*
In our algorithm, the jobs are sorted in ascending order of their burst time to give better turnaround time and waiting time like SRTN Algorithm. Performance of RR algorithm solely depends upon the size of time quantum. If it is very small, it causes too many context switches. If it is very large, the algorithm degenerates to FCFS. Taking into account the "80% criteria", our algorithm employs the use of dynamic quantum and also multi calculation of time quantum in a single cycle of round robin.

*3.2. Our Proposed Algorithm*

In our algorithm, when processes are already present in the ready queue, their arrival times are assigned to zero before they are allocated to the CPU. The burst time and the number of processes (n) are accepted as input. Let TQ be the time quantum. i and other integers specified are either counters or flag bits. The CST (Context Switching Time) in the algorithm is machine dependent. Hence the CPU Utilization depends on the register memory speed.

```
//Sort the processes in ascending order of their burst time
n → number of processes
 i, → counter value = 0
 while (ready queue is ! = NULL)
 if new process P_i arrived then sort it according to the burst time and
//Check P_i status
if P_i.status = 0 , then assign new counter C_i for this process end if
end if
//Find new time quantum
      m= median^{th} process
      MTQ = Median ( remaining burst time of all processes in ready queue )
      UTQ= Upper Quartile (n-m) (burst time of all remaining processes in ready queue after calculating the median)
//Continue executing the processes using MTQ and UTQ
for i =  1 to n loop
if ( i < m )
//Assign CPU to process P_i and give it time of slice= MTQ.
      P_i → MTQ_{cpu}
      i++;
   else
//Assign CPU to process P_i and give it time of slice= UTQ.
      P_i → UTQ_{cpu}
If P_i terminated normally and P_i.status = 0 then save P_i(C_i)
Let Pi.status = 1
  end if
 end for
end while
 Average waiting time, average turnaround time and number of context switches, Throughput, CPU Utilization (depending on CST) and CRITERIA % are calculated.
 End
```

Fig 2: Pseudo code for MDTQRR Algorithm

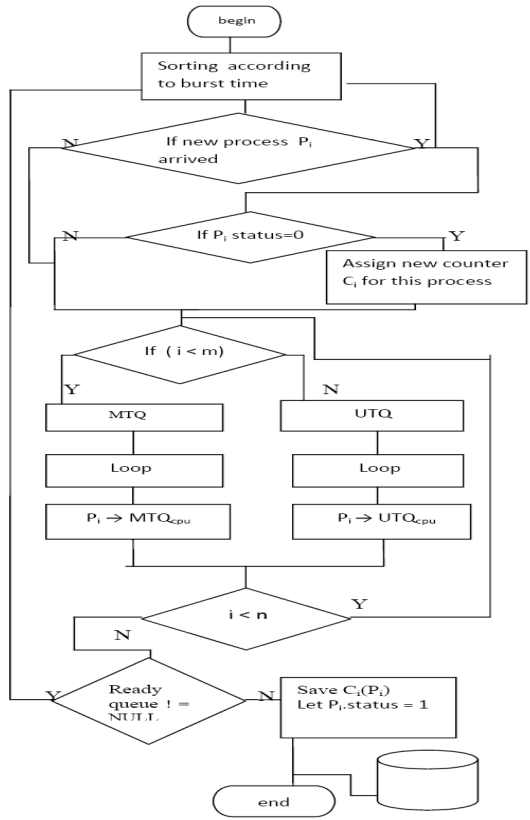

Fig 3: Flowchart for MDTQRR Algorithm

### 3.3. Time Complexity of MDTQRR Algorithm

The MDTQRR algorithm would be maintaining all jobs that are ready for execution in a queue. Each job insertion will be achieved in O(1), but job selection (to run next) and its deletion would require O(n) time, where n is the number of processes in the queue. MDTQRR simply maintaining all ready tasks in a sorted priority queue that will be used a linked list data structure. When a task arrives, a record for it can be inserted into the linked list in O(n) time where *n* is the total number of processes in the priority queue. Therefore, the time complexity of MDTQRR is equal to that of a typical linear sorting algorithm which is O(n). In our approach, whenever a task arrives, it is sorted according to its burst time in ascending manner and then executed .If a new task arrives it is then sorted with the remaining processes and then executed in the same way. To find a task with the lowest burst time the scheduler needs to search in the ready queue, then the order of searching would be O(n).

### 3.4. Illustration

Given the burst time sequence: 54 99 5 27 32. Initially the burst time of all the processes were sorted in ascending order which resulted in sequence 5 27 32 54 99. Then the *median* of the above burst time which was calculated to be 32 (MTQ) was assigned as the time quantum up to the median position of the processes. In the next step burst time for the rest processes are calculated by applying *upper quartile* method and it is found to be 99 (UTQ) only because after the last process no other process are there. When a process completes its burst time, it gets deleted from the ready queue automatically. So in this case, the processes P1, P2 and P3 were deleted from the ready queue, then P3 and P5 was given 99 as the time quantum so that it completes its execution. If the average time quantum is calculated using the MTQ and UTQ and compared with that calculated by using the CRITERIA percentage, we find that the algorithm is close to the 80% criteria. The above process was continued till all the processes were deleted from the ready queue. After the job scheduling is done for a given number of jobs, the THROUGHPUT is calculated. Since the cases are ideal, CST=0. Hence, CPU_TIME=TOTAL_TIME. Hence, going by the formulae, CPU UTILIZATION=100%.

## 4. Experimental Analysis

### 4.1. Assumptions

The environment where all the experiments are performed is a single processor environment and all the processes are independent. Time slice is assumed to be not more than the maximum burst time. All the attributes like burst time, number of processes and the time slice of all the processes are known before submitting the processes to the processor. All processes are CPU bound. No processes are I/O bound. Also, a large number of processes is assumed in the ready queue for better efficiency. Since, the cases are assumed to be close to ideal, the Context Switching Time is equal to zero i.e. there is no Context Switch Overhead incurred in transferring from one job to another.

### 4.2. Experimental Frame Work

Our experiment consists of several input and output parameters. The input parameters consist of burst time, arrival time, time quantum and the number of processes. The output parameters consist of average waiting time, average turnaround time, number of context switches, Throughput and Cpu Utilization.

### 4.3. Data set

We have performed three experiments for evaluating performance of our new proposed algorithm. In the above three cases experiments were performed by considering data set with different arrival time for each process.

### 4.4. Performance Parameters

The significance of our performance metrics for
experimental analysis is as follows:
*1) Turnaround time (TAT)*: For the better performance of the algorithm, average turnaround time should be less.
*2) Waiting time (WT)*: For the better performance of the algorithm, average waiting time should be less.
*3) Number of Context Switches (CS)*: For the better performance of the algorithm, algorithm, the number of context switches should be less.
*4) Throughput:* It is the number of processes completed per unit time.

### 4.5. Experiments Performed

To evaluate the performance of our proposed algorithm, we have taken a set of seven processes in six different cases. For simplicity, we have taken 7 processes. The algorithm works effectively for a very large number of processes. In each case, we have compared the experimental results of round robin scheduling algorithm with fixed time quantum Q with our proposed algorithm MDTQRR with dynamic time quanta MTQ and UTQ. Here we have assumed a constant time quantum Q equal to 40 in all the cases for RR, dynamic time quantum TQ and MTQ calculated by the median formula and the second dynamic time quantum UTQ calculated by the Upper Quartile formula. The CPU UTILIZATION is 100% in all cases, since CST =0 as assumed in ideal cases.

**Case 1:** We Assume five processes arriving at different times 0, 2, 5, 7, 9 respectively with increasing burst time (P1 = 10, P2 = 22, P 3 = 48, P 4 = 70, p5= 74) as shown in Table-1(upper). The Table-1(lower) shows the output using RR , SRBRR and MDTQRR algorithms. Figure-5, Figure-6 and Figure-7 shows Gantt chart for the algorithms respectively.

Table 1: Comparison between RR algorithm and our new proposed algorithm (case 1).

| Processes | Arrival Time | Burst Time |
|---|---|---|
| P1 | 0 | 10 |
| P2 | 2 | 22 |
| P3 | 5 | 48 |
| P4 | 7 | 70 |
| P5 | 9 | 74 |

| Algorithm | Time Quantum | Avg TAT | Avg WT | CS | Throug-hput |
|---|---|---|---|---|---|
| RR | 25 | 114.6 | 69.8 | 9 | 0.02 |
| SRBRR | 10,59,13,2 | 106.4 | 61.6 | 7 | 0.02 |
| MDTQRR | 10,59,74 | 94.6 | 50.2 | 4 | 0.02 |

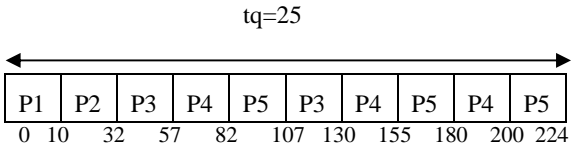

Fig.5: Gantt chart for RR in Table 1(case 1).

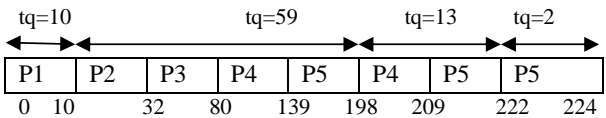

Fig.6: Gantt chart for SRBRR in Table 1 (case 1).

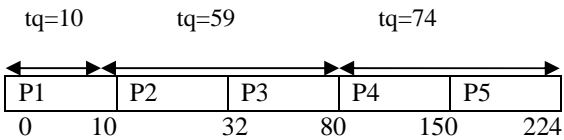

Fig.7: Gantt chart for MDTQRR in Table 1 (case 1).

**Case 2:** We Assume five processes arriving at different times 0, 6, 13, 21, 75 respectively with decreasing burst time (P1 = 73, P2 = 50, P3 = 23, P4 = 19, p5= 5) as shown in Table-2. The Table-2(lower) shows the output using RR , SRBRR and MDTQRR algorithms . Figure-9, Figure-10 and Figure-11 shows Gantt chart for the algorithms respectively.

Table 2: Comparison between RR algorithm and our new proposed algorithm (case 2).

| Processes | Arrival Time | Burst Time |
|---|---|---|
| P1 | 0 | 73 |
| P2 | 6 | 50 |
| P3 | 13 | 23 |
| P4 | 21 | 19 |
| P5 | 75 | 5 |

| Algorithm | Time Quantum | Avg TAT | Avg WT | CS | Throughput |
|---|---|---|---|---|---|
| RR | 25 | 101.8 | 67.8 | 7 | 0.03 |
| SRBRR | 73,23,23,27 | 87.4 | 53.4 | 5 | 0.03 |
| MDTQRR | 73,19,23,50 | 87.4 | 53.4 | 4 | 0.03 |

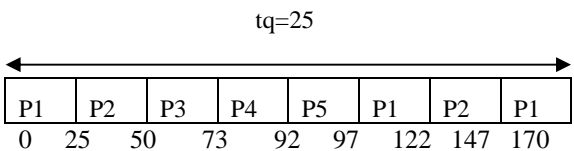

Fig.9: Gantt chart for RR in Table (case 2).

H. S. Behera et al. / Indian Journal of Computer Science and Engineering (IJCSE)

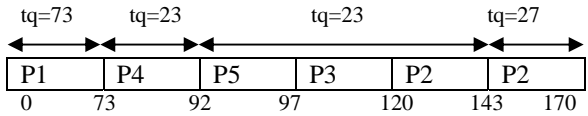

Fig.10: Gantt chart for SRBRR in Table 2 (case 2).

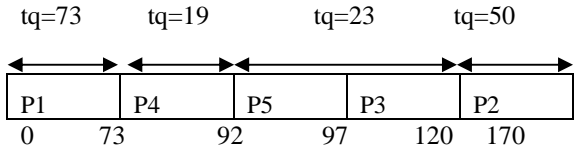

Fig.11: Gantt chart for MDTQRR in Table 2 (case 2).

**Case 3:** We Assume five processes arriving at different times 0, 6, 8, 9, 10 respectively with random burst time (P1 = 7, P2 = 15, P3 = 90, P4 = 42, p5= 8) as shown in Table-3(upper). The Table-3(lower) shows the output using RR , SRBRR and MDTQRR algorithms. Figure-13, Figure-14 and Figure-15 shows Gantt chart for the algorithms respectively.

Table 3: Comparison between RR algorithm and our new proposed algorithm (case 3).

| Processes | Arrival Time | Burst Time |
|---|---|---|
| P1 | 0 | 7 |
| P2 | 6 | 15 |
| P3 | 8 | 90 |
| P4 | 9 | 42 |
| P5 | 10 | 8 |

| Algorithm | Time Quantum | Avg TAT | Avg WT | CS | Throughput |
|---|---|---|---|---|---|
| RR | 25 | 72 | 39.6 | 8 | 0.03 |
| SRBRR | 7,15,42,48 | 52 | 19.6 | 5 | 0.03 |
| MDTQRR | 7,15,42,90 | 52 | 19.6 | 4 | 0.03 |

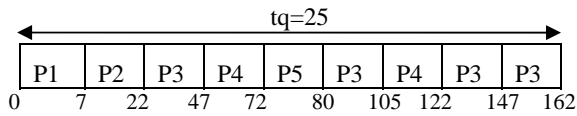

Fig.13: Gantt chart for RR in Table 3(case 3).

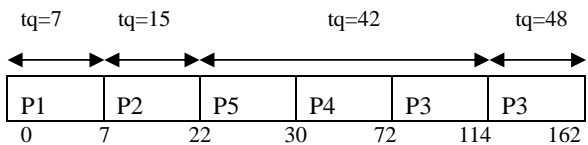

Fig.14: Gantt chart for SRBRR in Table 3(case 3).

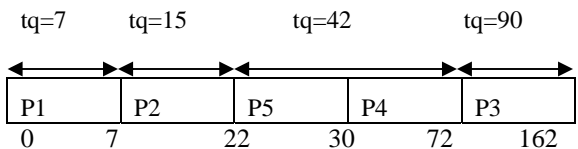

Fig.15: Gantt chart for MDTQRR in Table 3(case 3).

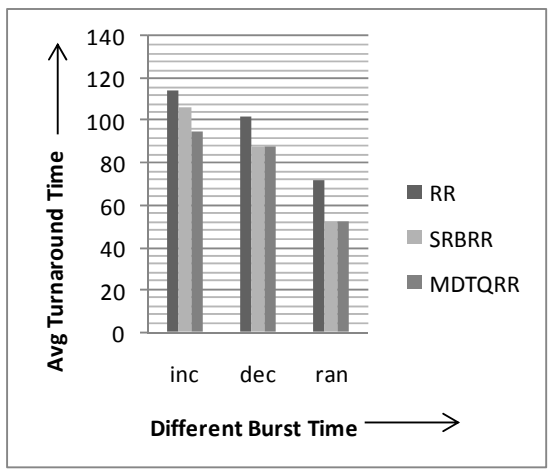

Fig.16: Comparison of average turnaround time RR, SRBRR and MDTQRR for increasing, decreasing and random burst sequence by taking arrival time into consideration.

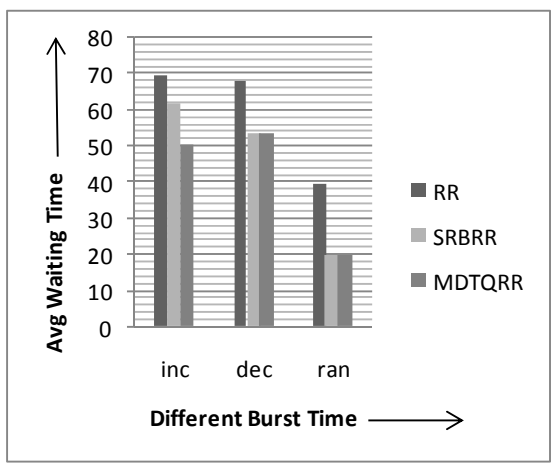

Fig.17: Comparison of average waiting time taking RR, SRBRR and MDTQRR for increasing, decreasing and random burst sequence by taking arrival time into consideration.

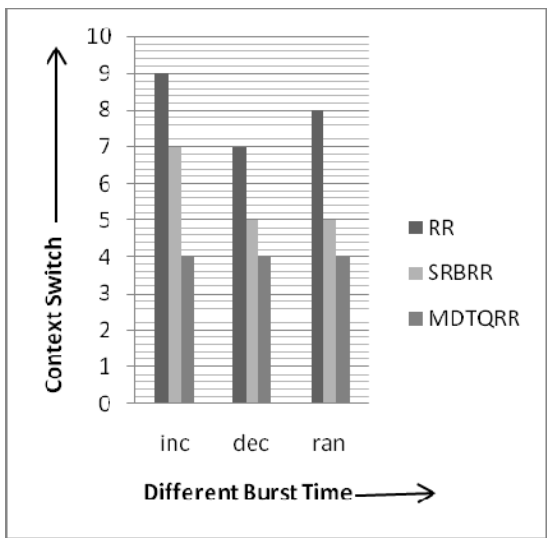

Fig.18: Comparison of context switches taking RR, SRBRR and MDTQRR for increasing, decreasing and random burst sequence by taking arrival time in consideration.